\documentclass[a4paper,11pt]{article}
\usepackage{amsfonts,amsmath,amssymb,bm,epsfig,url}
\usepackage[dvips]{color}

\usepackage{dcolumn}
\usepackage[pagebackref=false, colorlinks=true]{hyperref}
\usepackage{textcomp}
\usepackage{float}
\usepackage{subfigure}
\usepackage{graphicx}
\definecolor{redish}{rgb}{0.7,0.2,0.0}  
\definecolor{bluish}{rgb}{0.2,0.5,0.8}

\hypersetup{linkcolor=redish,          
                  citecolor=blue,        
                  filecolor=magenta,      
                  urlcolor=bluish}          

\DeclareFontFamily{U}{rsfs}{}         
\DeclareFontShape{U}{rsfs}{m}{n}{<5> rsfs5 <6><7> rsfs7          %
  <8><9><10><10.95><12><14.4><17.28><20.74><24.88> rsfs10}{}     %
\DeclareMathAlphabet{\mathfs}{U}{rsfs}{m}{n}

\def \O{\Omega}

\def \f{\frac}

\def \o{\omega}

\def \a{\alpha}

\def \p{\partial}

\def \a{\alpha}

\def \th{\theta}

\title{Behavior of a test gyroscope moving 
towards a rotating traversable wormhole}
\author{Chandrachur Chakraborty\footnote{chandrachur.chakraborty@tifr.res.in}
\\{\it Tata Institute of Fundamental Research, Mumbai 400005, India}
\\
{}
\\
Parthapratim Pradhan\footnote{pppradhan77@gmail.com}
\\{\it Department of Physics, Vivekananda Satabarshiki Mahavidyalaya,
West Midnapur 721513 India}
}
\date{}

\begin{document}

\maketitle

\begin{abstract}
The geodesic structure of the Teo wormhole is briefly discussed
and some observables are derived that promise to be of use in detecting a
rotating traversable wormhole indirectly, if it does exist.
We also deduce the exact Lense-Thirring (LT) precession frequency of a test 
gyroscope moving toward a rotating traversable Teo wormhole. 
The precession frequency diverges on the ergoregion, a behavior intimately related to and governed by the geometry 
of the ergoregion, analogous to the situation in a Kerr spacetime. Interestingly,
it turns out that here the LT precession is inversely proportional to the angular momentum ($a$) of the 
wormhole along the pole and around it in the strong gravity regime, 
a behavior contrasting with its direct variation with $a$ in the case of other compact objects.
In fact, divergence of LT precession inside the ergoregion 
can also be avoided if the gyro moves with a non-zero 
angular velocity in a certain range. As a result, the spin precession frequency of the gyro can be made finite throughout its 
whole path, even very close to the throat, during its travel to the wormhole. Furthermore, 
it is evident from our formulation
that this spin precession not only arises due to curvature 
or rotation of the spacetime but also
due to the non-zero angular velocity of the spin when it does not move along a
geodesic in the strong gravity regime.  
If in the future, interstellar travel indeed becomes possible through a wormhole 
or at least in its vicinity, our results would prove useful
in determining the behavior of a test gyroscope which is known to serve  
as a fundamental navigation device.
\end{abstract}


\maketitle

\section{Introduction}
A wormhole or ``Einstein-Rosen bridge'' is a hypothetical
topological feature which would be a shortcut linking two 
separate points in spacetime. In 1916, Flamm first showed that there can
be a tunnel between distant parts
of space \cite{fl} and such a tunnel is described by the spatial part of
the Schwarzschild metric, taken separately from time. Later,
Einstein and Rosen also proposed solutions which involve the mathematical
representation of the physical space by a space of two identical sheets,
a particle being represented by a `bridge' connecting these sheets \cite{er}.
So, the wormhole is basically a ``tunnel'' with two
ends and it may connect extremely long distances (say, billions of light years 
or a part of the Universe with another part sufficiently remote) in a 
very very short distances (say, within a few meters).
The term `wormhole' was first coined by Wheeler in 1957 and he also
proposed the concept of the charge-carrying microscopic wormholes \cite{w}.
However, the possibility of traversable wormhole was first demonstrated by
Ellis \cite{hg} in 1973 and independently by Bronnikov \cite{bro} in the same year.
A duplicate of the Ellis wormholes was presented by Morris and Thorne \cite{mt} 
just as `a tool for teaching general relativity' and they made some 
important observations. For example, they showed
that to stabilize a traversable wormhole, i.e., to hold a wormhole open 
permanently, one needs exotic matter which must violate the 
weak energy condition \cite{mty}. Later, Visser \cite{v} developed some
interesting examples of traversable wormholes to minimize the use of exotic
matter and showed that it was possible for a traveler to
traverse such a wormhole without passing through a region of exotic matter.
Here, we should note that a class of exact stationary and axisymmetric 
multi-wormhole solutions of the Einstein-Maxwell scalar field equations
was generated by Cl\'ement \cite{cl1,cl2,cl3} and it was shown that these rotating 
wormhole solutions were asymptotically NUT-like.

Though, wormholes have not been observationally detected till
now, Zhou et al. \cite{zhou} have recently investigated  possible observational
signatures to identify the astrophysical rotating Ellis wormhole
which is basically a massive and compact object. Studying 
the iron line profile in the X-ray
reflected spectrum of a thin accretion disk around the rotating Ellis wormhole,
they have found some specific observational signatures which can be used to distinguish
the wormhole from the Kerr black hole. Thus, it may be possible to find the wormhole
in the near future. In this regard, we calculate some specific
observables for Teo wormhole related to the timelike and lightlike geodesics
(as the geodesics in Teo wormhole have already been 
studied in Refs.\cite{tb,tb2} during the calculation of high energy 
particle collision and collisional Penrose process in Teo wormhole,
we do not want to repeat it here), which
may help us to distinguish a rotating traversable wormhole from
other compact objects in future, if it in fact exists. Moreover, 
these observables are also necessary for completeness of our study
of the spin precession behavior in Teo wormhole. We will 
discuss these issues as we proceed.

In this article, we explore  how the spin of 
a gyroscope, a fundamental navigation device, attached to a spaceship, 
would behave in the vicinity of a wormhole if interstellar travel
is possible in future through a traversable wormhole. For our analysis, we consider here
the {\it simplest possible} rotating traversable wormhole which has recently been 
discovered by Teo \cite{teo}. From our analysis the following interesting physical
questions could be answered : (i) how will the gyroscope of an astronaut holding 
his/her spaceship at a constant distance from a wormhole behave? (ii) will the
astronaut see any change in behavior of the gyro precession, if 
the spaceship rotates with some finite angular velocity around the rotating 
wormhole? (iii)(if the spaceship moves toward an
unknown object) can the astronaut distinguish the wormhole from 
other compact objects, using the behavior of the spin precession of the gyroscope?

To answer the first question, i.e., what an astronaut 
will see the behavior of the gyro attached to his/her spaceship to be,
we have to calculate the Lense-Thirring (LT) precession in Teo wormhole, using
the general formulation of LT effect derived in Ref.\cite{cm}. We show 
an `adverse effect' of LT precession in the wormhole spacetime, which can 
be used to distinguish a rotating traversable wormhole from other 
compact objects.
A more general formulation of the spin precession of a test gyro
has recently been derived in Ref.\cite{cpk}, which is more realistic
in this sense that the gyro (attached to the spaceship) can move
towards the wormhole with an arbitrary angular velocity in a geodesic or
non-geodesic path. This will give us  answers to the second question
as well as partly to the third question. 

The paper is organized as follows : in Sec.\ref{cpow}, we briefly discuss
the timelike and lightlike geodesics in the rotating
traversable Teo wormhole and work out the radius
of the circular photon orbit (CPO). As an integral part of the study of
geodesics, Sec.\ref{epi} has been devoted to 
compute the three fundamental frequencies (Kepler frequency, radial and 
vertical epicyclic frequencies) of a test particle which rotates around
the Teo wormhole and using this, we derive the radii of the 
innermost-stable-circular-orbits (ISCOs). In Sec.\ref{sec4}, we derive the 
exact LT precession frequency of a
test gyro which {\it moves} toward the said wormhole. Using the general formulation
of the spin precession \cite{cpk}, we find the behavior of a test gyro 
(in Sec. \ref{genf}) which rotates
with a non-zero angular velocity around the Teo wormhole or travels toward it.
Finally, we conclude in Sec.\ref{dis}.

\section{\label{cpow}Lorentzian traversable wormhole and Circular Photon Orbit}
In this section, we briefly describe the geodesic structure of the traversable 
wormhole. For this, we first write the metric of the traversable wormhole as \cite{teo} :
\begin{eqnarray}
 ds^2 &=& -N^2dt^2+\left(1-\frac{b}{r}\right)^{-1}dr^2+r^2K^2\left[d\theta^2+\sin^2\theta(d\phi-\omega dt)^2\right]
\label{w1}
\end{eqnarray}
where  $N,\omega, K, b$ are functions of  $(r,\theta$). This metric is regular on the
symmetry axis $\th=0, \pi$ and it indicates 
two identical, asymptotically flat spacetimes which are connected together 
at the throat, i.e., $r=b>0$. The discriminant of the metric 
$D=g_{t\phi}^2-g_{tt}g_{t\phi}=N^2K^2r^2 \sin^2\th$ indicates that the event
horizon can exist for $N=0$. Since $N$ is finite and also $N\neq 0$ in the 
case of a wormhole, the event horizon does not exist and the spacetime contains 
no curvature singularity at all. 
According to Teo, we can take the parameters $N$ and $\o$ for a particular
traversable wormhole as follows \cite{teo}:
\begin{eqnarray}
N=K=1+\frac{16a^2 d~ \cos^2\theta}{r} \,\,\,\, , \,\,\,
\o=\frac{2a}{r^3}  ~\label{w2}
\end{eqnarray}
where $b$ and $d$ \cite{tb} are positive constants and $a$ is 
the total angular momentum of the wormhole. The throat  of
the wormhole occurs at $r=b$ and it looks like a
`peanut-shell' (see FIG.1 of Ref.\cite{teo}). However, the radial coordinate
always satisfies the following condition : $r\geq b$ to maintain the 
structure of the wormhole spacetime. Therefore, as $b$ determines the spatial
shape of the wormhole, $b$ is also called the shape function.
For a fast-rotating wormhole $a>\frac{b^2}{2}$ and the ergoregion occurs in
the range of $b^2<r^2 \leq |2a \sin\th|$. Interestingly, the ergoregion 
does not extend to the poles (i.e., $\th=0$ and $\th=\pi$), rather it forms a 
tube like structure around the equatorial plane.

Here, we are mainly interested to derive the radius of CPO 
in the equatorial plane of the Teo wormhole. Therefore, the metric (Eq. \ref{w1})
reduces to the following form :
\begin{eqnarray}
 ds^2 &=& -\left(1-\frac{4a^2}{r^4}\right)dt^2-\frac{4a}{r}dt d\phi
 +\left(1-\frac{b}{r}\right)^{-1}dr^2+r^2d\phi^2 .
\label{w3}
\end{eqnarray}

To determine the geodesics in the equatorial plane, we follow 
Refs.\cite{sch,cqg} and compute the geodesic motion of a neutral 
test particle setting $\dot{\theta}=0$ and $\theta=\frac{\pi}{2}$.
Therefore, the necessary Lagrangian for the geodesic motion should be
\begin{eqnarray}
2{\cal L} = -\left(1-\frac{4a^2}{r^4}\right)\,{\dot{t}}^2
-\frac{4a}{r}\,\dot{t}\,\dot{\phi}+
\left(1-\frac{b}{r}\right)^{-1}\,{\dot{r}}^2+r^2\,{\dot{\phi}}^2 ~\label{w4} .
\end{eqnarray}
The generalized momenta can be written as
\begin{eqnarray}
p_{t} &=& -\left(1-\frac{4a^2}{r^4}\right) \,\,\dot{t}
-\frac{2a}{r} \,\dot{\phi}=-{\cal E} ~,\label{w5}
\end{eqnarray}
\begin{eqnarray}
p_{\phi} &=& -\frac{2a}{r}\,\dot{t} +r^2\,\dot{\phi}=\ell ~,\label{w6}
\end{eqnarray}
and
\begin{eqnarray}
 p_{r} &=& \left(1-\frac{b}{r}\right)^{-1}\, \dot{r}  ~.\label{w7}
\end{eqnarray}
Here, $\dot{t},~\dot{r}$ and $\dot{\phi}$ indicate the differentiation with respect 
to the affine parameter $\tau$. Since the metric
does not depend on `$t$' and `$\phi$', therefore $p_{t}$ and $p_{\phi}$ are conserved 
quantities. The independence of the metric
on `$t$' and `$\phi$' manifests the stationary and axially symmetric 
character of the wormhole space-time. However, solving Eq. (\ref{w5}) 
and Eq. (\ref{w6}) for $\dot{t}$ and $\dot{\phi}$ we obtain
\begin{eqnarray}
\dot{t} &=& \frac{1}{r^2}\left[{\cal E}r^2 -\frac{2a\ell}{r}\right]  ~\label{w8}
\end{eqnarray}
and
\begin{eqnarray}
 \dot{\phi} &=& \frac{1}{r^2}\left[\frac{2a {\cal E}}{r}
+\ell\left(1-\frac{4a^2}{r^4}\right)\right]   ~\label{w9},
\end{eqnarray}
where  ${\cal E}$ and $\ell$ are the energy and angular momentum per 
unit mass of the test particle, respectively. The normalization condition of the four 
velocity ($u^{\mu}$) gives another integral equation for the geodesic motion,
which can be written as:
\begin{eqnarray}
g_{\mu\nu}u^{\mu}u^{\nu} &=& \epsilon ~\label{w10}
\end{eqnarray}
or
\begin{eqnarray}
-{\cal E}\dot{t}+\ell \dot{\phi}+\frac{{\dot{r}}^2}{\left(1-\frac{b}{r}\right)}&=& \epsilon ~\label{w11}
\end{eqnarray}
where  $\epsilon=-1$ for time-like geodesics, $\epsilon=0$ for light-like geodesics and
$\epsilon=+1$ for space-like geodesics.
Using Eq.(\ref{w8}) and Eq.(\ref{w9}), we eliminate $\dot{t}$ and $\dot{\phi}$
from Eq. (\ref{w11}) and finally we obtain the radial equation
that governs the geodesic motion of the test particle in the Teo wormhole as :
\begin{eqnarray}
\dot{r}^{2} & =& \left(1-\frac{b}{r}\right)\left[{\cal E}^2-\frac{4a\ell {\cal E}}{r^3}-
\frac{\ell^2}{r^2}+\frac{4a^2\ell^2}{r^6} + \epsilon \right] ~.\label{w12}
\end{eqnarray}
For light-like geodesics, we set $\epsilon=0$ and hence, Eq. (\ref{w12}) reduces to :
\begin{eqnarray}
\dot{r}^{2} &=& \left(1-\frac{b}{r}\right)\left[{\cal E}^2-
\frac{4a\ell {\cal E}}{r^3}-\frac{\ell^2}{r^2}+\frac{4a^2\ell^2}{r^6}
\right]~.\label{w13}
\end{eqnarray}
We introduce an impact parameter $\eta_c=\frac{\ell_c}{{{\cal E}_c}}$ to determine
the radius ($r_{c}$) of the unstable CPO for
${\cal E}={\cal E}_{c}$ and $\ell={\ell}_{c}$. Therefore, we obtain two 
important equations from Eq.(\ref{w13}), which can be written as
\begin{eqnarray}
r_{c}^{6}-\eta_{c}^2 r_{c}^4-4a \eta_{c} r_{c}^3+4a^2 \eta_{c}^2 &=& 0 
\label{w14}
\end{eqnarray}
and
\begin{eqnarray}
3 r_{c}^3-2 \eta_{c}^2r_{c}-6 a \eta_{c} &=& 0 ~.\label{w15}
\end{eqnarray}
Now, solving for $\eta_c$, we obtain from Eq.(\ref{w15}) :
\begin{eqnarray}
\eta_{c} &=& \frac{-3a\pm \sqrt{6r_{c}^4+9a^2}}{2r_{c}}~.\label{w16}
\end{eqnarray}
Here $`+$' and $`-$' signs appear for the direct and retrograde orbits
respectively. Now, inserting Eq. (\ref{w16}) in (\ref{w14}), we find the 
expression to determine the radius of the CPO as,
\begin{eqnarray}
r_{c}^{8}-15a^2r_{c}^4\pm \left(ar_{c}^4+12a^3\right)\sqrt{6r_{c}^4+9a^2}-36a^4 = 0  
.\label{w17}
\end{eqnarray}
Eliminating the square root from the above equation we obtain
\begin{eqnarray}
r_{c}^{12}-36 a^2 r_{c}^8 -1050 a^6 &=& 0  ~\label{w18}
\end{eqnarray}
and solving the above equation, the radius of the CPO is determined as
\begin{eqnarray}
 r_{c} \approx \pm 2.4626~\sqrt{a} ~
 \label{cpo}
\end{eqnarray}
where the upper sign is applied to the direct orbits and the lower 
one is applied to the retrograde orbits. Eq.(\ref{cpo}) is important for 
us as it will be applied in Sec.\ref{gem} to calculate the effect of 
frame-dragging at the CPO.

\section{\label{epi} Observables in the rotating traversable wormhole spacetime}

It has already been stated that Zhou et al. \cite{zhou} have recently studied the X-ray
reflected spectrum of a thin accretion disk around the rotating Ellis wormhole which
is a massive and compact object. They have proposed that the wormholes may look like black holes
and they have found some specific observational signatures by which it is possible 
to distinguish rotating wormholes from Kerr black holes. 
The Teo traversable wormhole being considered in this article,
may not be a compact object but if a test particle 
moves in this wormhole spacetime it experiences some change in its periodic motion
along the  co-ordinates  $r$, $\th$ and $\phi$. Here, it is necessary to study 
these periodic variation due to not only the completeness of our study of the geodesic motion
of the test particle but also because it helps us to derive the orbital plane precession
frequency and the periastron precession frequency which is an integral part of 
the study of geodesics. If it is possible to detect the wormhole in near future 
and if we want to conduct realistic experiments in the rotating wormhole spacetime then
these precession frequencies will serve as  observables. The periodic motion 
along $r$ and $\th$ of a test particle is known as radial epicyclic frequency 
and vertical epicyclic frequency respectively. The periodic variation along $\phi$
is very well-known to us, it is the Kepler frequency.

At first, we consider a general stationary and axisymmetric spacetime as follows:
\begin{eqnarray}
 ds^2=g_{tt}dt^2+2g_{t\phi}d\phi dt+g_{\phi\phi}d\phi^2+g_{rr}dr^2+g_{\theta\theta}d\theta^2
\end{eqnarray}
where $g_{\mu\nu}=g_{\mu\nu}(r, \theta)$. In this spacetime, 
the proper angular momentum ($l$) of a test particle can be defined as : 
\begin{eqnarray}
 l=-\frac{g_{t\phi}+\Omega_{\phi} g_{\phi\phi}}{g_{tt}+\Omega_{\phi} g_{t\phi}}
\end{eqnarray}
where, $\Omega_{\phi}$ is the Kepler frequency of the test particle which
is defined as \cite{don}
\begin{eqnarray}
 \Omega_{\phi}=\frac{d\phi/d\tau}{dt/d\tau}=\frac{d\phi}{dt}
 =\frac{-g_{t\phi}'\pm \sqrt{g_{t\phi}'^2-g_{tt}'g_{\phi\phi}'}}{g_{\phi\phi}'}\mid_{r=constant , \theta=constant} .
\end{eqnarray}
where the prime denotes the partial differentiation with respect to $r$.

For Teo rotating traversable wormhole (Eq.(\ref{w3})) we can calculate the 
Kepler frequency which comes out as, 
\begin{eqnarray}
 \Omega_{\phi}^{d}=\frac{2a}{r^3} \label{kd}
\end{eqnarray}
and
\begin{eqnarray}
 \Omega_{\phi}^{g}=-\frac{4a}{r^3}
 \label{kg}.
\end{eqnarray}
where the negative sign indicates that the rotation is in the reverse direction. Suffixes $d$ and
$g$ stand for the direct and retrograde rotation respectively.

The general expressions for calculating the radial ($\O_r$) and vertical
($\O_{\th}$) epicyclic frequencies are \cite{don}
\begin{eqnarray}\nonumber
 &&\O_r^2=\f{(g_{tt}+\O_{\phi}g_{t\phi})^2}{2~g_{rr}}~\p_r^2~U
 \\
&=&\f{(g_{tt}+\O_{\phi}g_{t\phi})^2}{2~g_{rr}}\left[\p_r^2\left({g_{\phi\phi}}/{Y}\right)
 +2l~\p_r^2\left({g_{t\phi}}/{Y}\right)+l^2~\p_r^2\left({g_{tt}}/{Y}\right) \right]|_{r=const., \th=const.}\nonumber
 \label{re}
 \\
\end{eqnarray}
and
\begin{eqnarray}\nonumber
 &&\O_{\th}^2=\f{(g_{tt}+\O_{\phi}g_{t\phi})^2}{2~g_{\th\th}}~\p_{\th}^2~U
 \\
 &=&\f{(g_{tt}+\O_{\phi}g_{t\phi})^2}{2~g_{\th\th}}\left[\p_{\th}^2\left({g_{\phi\phi}}/{Y}\right)
 +2l~\p_{\th}^2\left({g_{t\phi}}/{Y}\right)+l^2~\p_{\th}^2\left({g_{tt}}/{Y}\right) \right]|_{r=const., \th=const.}\nonumber
 \label{ve}
 \\
\end{eqnarray}
respectively and $Y$ can be defined as
\begin{eqnarray}
 Y=g_{tt}g_{\phi\phi}-g_{t\phi}^2 .
\end{eqnarray}
In our case, we are confined to $\th=\pi/2$ which is physically reliable. Thus, we 
obtain for the Teo wormhole
\begin{eqnarray}
 Y=-r^2 
\end{eqnarray}
for $\th=\pi/2$ and 
\begin{eqnarray}
 \p_r^2~U|_{r=r_{const}, \th=\pi/2}=\f{6l}{r^8} (lr^4-28a^2l+8ar^3).
\end{eqnarray}

As the proper angular momentum are calculated for this particular wormhole as
\begin{eqnarray}
 l^{d}=0
 \label{ld}
\end{eqnarray}
and
\begin{eqnarray}
 l^{g}=\frac{6ar^3}{12a^2-r^4} 
 \label{lg}
\end{eqnarray}
we obtain the radial epicyclic frequencies $(\O_r)$ from Eq.(\ref{re}) for 
the direct and retrograde rotation as 
\begin{eqnarray}
 \O_r^{2~(d)}=0
 \label{dre}
\end{eqnarray}
and
\begin{eqnarray}
\O_r^{2~(g)}=-\f{36a^2\left(1-\f{b}{r}\right)}{r^{10}}~(r^4+36a^2)
  \label{gre}
\end{eqnarray}
respectively.

\subsection{Radius of the innermost stable circular orbits (ISCOs)}

It is well-known that the square of the radial epicyclic frequency 
is equal to zero at the ISCO and it is negative for the smaller radius,
which shows the radial instabilities for orbits with radius smaller
than the ISCO. In our case, it is easily seen from
Eq.(\ref{gre}) that $\O_r^{2~(g)}$ vanishes at $r=b$ ($r=b$ is actually the radius
of the throat of the wormhole) but it becomes negative
for $r > b$. This means that only one stable circular orbit is possible
at $r=b$ for the retrograde rotation and the orbits occurred at $r > b$ are unstable.
Thus, ISCO is meaningless in this particular case. On the other hand,
Eq.(\ref{dre}) reveals that the radial epicyclic frequency 
vanishes for all direct orbits. This means that all direct orbits 
are stable in the Teo wormhole spacetime and no instability is there.
Thus, ISCO should occur at $r=b$ for the timelike
geodesics and it could be stated that the ISCO coincides at the throat
of the Teo wormhole in case of the prograde rotation. Here, we can note that 
the ISCO coincides at the throat \cite{zhou} for the direct orbit in case 
of a rotating Ellis wormhole also. 

\subsection{Periastron precession frequency}

It is very surprising that the radial epicyclic frequency 
vanishes for the direct orbit as the angular momentum is zero in this case.
Interestingly, the angular velocity
is non zero (Eq.(\ref{kd})) for the direct orbit, which can be an example of the zero 
angular momentum observer with non-zero angular velocity. Therefore, if a test
particle rotates in any circular geodesic (direct orbit) in this spacetime, it will
carry zero angular momentum with non-zero angular velocity.
Now, we can easily calculate the {\it periastron precession rate} $(\O_{per}^d)$
for the direct orbit which turns out to be
\begin{eqnarray}
 \O_{\rm per}^d=\O_{\phi}^d-\O_r^d=\O_{\phi}^d=\f{2a}{r^3}.
 \label{perd}
\end{eqnarray}
Thus, it is really remarkable that the periastron precession rate is equal 
to the Kepler frequency (Eq.\ref{kd})
for the direct orbit in the Teo wormhole. This means that the precession rate of the 
orbit ($\O_{\rm per}^d$) of the test particle is same as the angular velocity of it.
But, for the retrograde rotation the 
{\it periastron precession frequency} comes out as
\begin{eqnarray}
 \O_{\rm per}^g=\O_{\phi}^g-\sqrt{|\O_r^{2~(g)}|}=-\f{2a}{r^3}\left[2+\f{3}{r^2}\left(1-\f{b}{r}\right)^{\f{1}{2}}(r^4+36a^2)^{\f{1}{2}}\right].
\label{perg}
 \end{eqnarray}
Negative sign confirms that the rotation is in the reverse direction.

\subsection{Orbital plane precession frequency}

To calculate the orbital plane precession frequency, we first have to obtain
the vertical epicyclic frequency for the Teo wormhole and for that we calculate
\begin{eqnarray}
 \p_{\th}^2~U|_{r=r_{const}, \th=\pi/2}=\f{64a^2d}{r}-l\f{128a^3d}{r^4}+\f{2l^2}{r^2}
 \left[1-\f{32a^2d}{r}\left(1-\f{4a^2}{r^4}\right)\right].
\end{eqnarray}
Now, substituting it into Eq.(\ref{ve}) and also taking the values of $l$ from
Eq.(\ref{ld},\ref{lg}) we obtain
\begin{eqnarray}
 \O_{\th}^{2~(d)}=\f{32a^2d}{r^3}
 \label{vd}
\end{eqnarray}
for the direct orbit and for the retrograde orbit we obtain
\begin{eqnarray}
 \O_{\th}^{2~(g)}=\f{4a^2}{r^7}~[9r+8d(r^4-36a^2)].
 \label{vg}
\end{eqnarray}

As it has already been derived the exact 
Kepler frequencies and the vertical epicyclic frequencies of a test particle which rotates
in a circular orbit around the traversable wormhole in the equatorial plane, now we can obtain
the {\it nodal precession frequency} $(\O_{nod})$ easily. It is also called as the orbital plane precession 
frequency or the Lense-Thirring (LT) precession frequency of a {\it test particle}. 
Using Eqs.(\ref{kd},\ref{kg}) and Eqs.(\ref{vd},\ref{vg})
we obtain :
\begin{eqnarray}
 \O_{\rm nod}^d=\O_{\phi}^d-\O_{\th}^d=\f{2a}{r^3}(1-\sqrt{8r^3d})
 \label{dnod}
\end{eqnarray}
for the direct orbit and for the retrograde orbit it comes out as
\begin{eqnarray}
 \O_{\rm nod}^g=\O_{\phi}^g-\O_{\th}^g=-\f{2a}{r^3}\left[2+\left(9+\f{8d}{r}(r^4-36a^2)\right)^{\f{1}{2}}\right]
\label{gnod}
 \end{eqnarray}
where the negative sign indicates that the rotation is in the reverse direction. 

It is important to mention here that Eq.(\ref{perd}) and Eq.(\ref{dnod}) are physically meaningful
as the stable circular orbits exist for $r \geq b$ but Eq.(\ref{perg}) and Eq.(\ref{gnod})
are not important except that particular orbit occurred at $r=b$
because only this orbit is stable.

\section{\label{sec4}Lense-Thirring precession of a test gyro in the 
rotating traversable wormhole spacetime}
We are keen to derive the Lense-Thirring (LT) precession \cite{lt} frequency of a test gyro \cite{schiff} in a 
general rotating traversable wormhole spacetime, which could be regarded as a more 
realistic observable for the indirect detection of a wormhole. In 2004, NASA sent the Gravity 
Probe B (GP-B) satellite to measure the Geodetic and LT precession frequencies of a test gyro due to the 
rotation of the earth and the result had been published by Everitt et al. in 2011 \cite{ev}. 
It may play an important role in the future
astrophysical observational purpose to measure the LT precession frequency of a wormhole to distinguish its 
nature from the other compact objects as well. We note that the exact LT 
precession rate of a test gyro in the Kerr spacetime \cite{cm} and some other
axisymmetric spacetimes \cite{cp, cmb} have already been derived. 
Now, this formulation can also be applied to obtain the frame-dragging
effect in case of the rotating wormhole as well.
The canonical metric for a general stationary, axisymmetric traversable wormhole has been written 
in Eq.(\ref{w1}). Using this metric, the exact LT precession rate of a test gyro
relative to the Copernican system (or `fixed stars')
in the rotating traversable wormhole spacetime can be expressed in the orthonormal coordinate basis as:
\begin{eqnarray}\nonumber
&&\vec{\Omega}_{LT}=\frac{1}{2NK(-N^2+\omega^2r^2K^2\sin^2\theta)}.   
\\
&&\left[K\left(1-\frac{b}{r}\right)^{1/2}\sin\theta[N^2(Kr\omega_{,r}+2\omega rK_{,r}+2\omega K)
+\omega^2 r^3K^3\omega_{,r}\sin^2\theta-2\omega KrNN_{,r}]\hat{\theta}\right.
\label{ltnp} \nonumber
\\ 
&& \left.-[N^2(K\omega_{,\theta}\sin\theta+2\omega K_{,\theta}\sin\theta +2\omega K\cos\theta)
+\omega^2 r^2K^3\omega_{,\theta}\sin^3\theta-2\omega KNN_{,\theta}\sin\theta]\hat{r}\right]\nonumber
\\
\end{eqnarray}
and the modulus of the above LT precession rate is
\begin{eqnarray}\nonumber
&& \Omega_{LT}=|\vec{\Omega}_{LT}(r,\theta)| 
\\&=&\frac{1}{2NK(-N^2+\omega^2r^2K^2\sin^2\theta)}. \nonumber
\\
&&\left[K^2\left(1-\frac{b}{r}\right)\sin^2\theta[N^2(Kr\omega_{,r}+2\omega rK_{,r}+2\omega K)
+\omega^2 r^3K^3\omega_{,r}\sin^2\theta-2\omega KrNN_{,r}]^2\right. \nonumber
\\ 
&&
\left.+[N^2(K\omega_{,\theta}\sin\theta+2\omega K_{,\theta}\sin\theta +2\omega K\cos\theta)
+\omega^2 r^2K^3\omega_{,\theta}\sin^3\theta-2\omega KNN_{,\theta}\sin\theta]^2\right]^\frac{1}{2}.
\nonumber
\\
\end{eqnarray}

\subsection{\label{lt}LT precession of a test gyro in the Teo wormhole}
According to Teo, we can take $N$ and $\o$ as the following:
\begin{eqnarray}
 N=K=1+\frac{16a^2 d~ \cos^2\theta}{r} \,\,\,\, , \,\, \omega=\frac{2a}{r^3}
\end{eqnarray}
and can calculate the exact LT precession rate of a test gyro for the said 
wormhole, which can be expressed as 
\begin{eqnarray}
 \vec{\Omega}_{LT}=\frac{a}{Nr^3\left(1-\frac{4a^2}{r^4}\sin^2\theta \right)}.
\left[N\sin\theta\left(1-\frac{b}{r}\right)^{\frac{1}{2}}\left(1+\frac{12a^2}{r^4}\sin^2\theta \right)\hat{\theta}
+2\cos\theta \,\, \hat{r}\right] \nonumber .
\\
\label{sw}
\end{eqnarray}
Along the pole we can take $\theta=0$ and the LT precession frequency can be
obtained as
\begin{eqnarray}
 |\vec{\Omega}_{LT}|=\frac{2a}{r^2(r+16a^2 d)} .
 \label{sl}
\end{eqnarray}
Similarly, we can take $\theta=\pi/2$ along the equator and the LT precession 
frequency comes out as
\begin{eqnarray}
 |\vec{\Omega}_{LT}|=\frac{a}{r^3\left(1-\frac{4a^2}{r^4}\right)}. 
 \left(1-\frac{b}{r}\right)^{\frac{1}{2}}\left(1+\frac{12a^2}{r^4}\right)\hat{\theta}.
\label{se} 
 \end{eqnarray}
 
Weak field approximation $(r >> \sqrt{a})$ can be taken at a large distance, 
for which Eq.(\ref{sw}) reduces to
\begin{eqnarray}
 \vec{\Omega}_{LT}=\frac{a}{Nr^3}.
\left[N\sin\theta\left(1-\frac{b}{r}\right)^{\frac{1}{2}}\hat{\theta}
+2\cos\theta \,\, \hat{r}\right] .
\label{sw1}
\end{eqnarray}
Here, we should note that the LT precession formulation is
valid only outside the ergoregion \cite{cm} as an observer cannot remain
stationary with zero angular velocity. Thus, 
Eq.(\ref{sw}) and the related equations will valid only in the following range:
\begin{eqnarray}
 r \geq b \,\,\,\,\, \mbox{and} \,\,\,\,\, r^2 > 2a\sin\th  
\end{eqnarray}
and both of the above conditions must be hold. As we have stated earlier that the 
ergoregion ($r_e$) occurs for the rotating Teo
wormhole in this range $b < r_e \leq \sqrt{2a\sin\th}$, the calculation of LT precession 
frequency is valid only for $r>\sqrt{2a\sin\th}$. It should be noticed that
the ergoregion does not fully surround the throat but forms a tube around the equatorial
region of the said wormhole. Unlike a rotating black hole spacetime, the ergoregion 
of a rotating wormhole
does not coincide with the horizon at the pole as the event horizon does not
exist in this spacetime at all. At the equatorial plane, the ergoregion is 
extended upto $r_e=\sqrt{2a}$ and it decreases slowly if we move to the pole from
the equator. Finally, the ergoregion vanishes $(r_e=0)$ at the pole for $\th=0$.

\begin{figure}[]
 \begin{center}
\subfigure[At $0^0$ (along the pole)]{
\includegraphics[width=2.3in,angle=0]{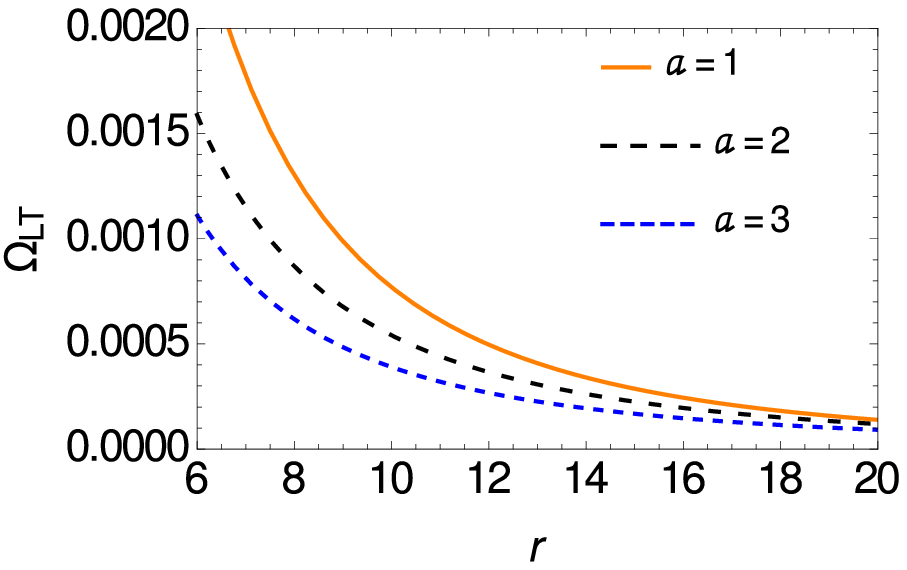}} 
 \subfigure[At $12^0$ ]{
 \includegraphics[width=2.3in,angle=0]{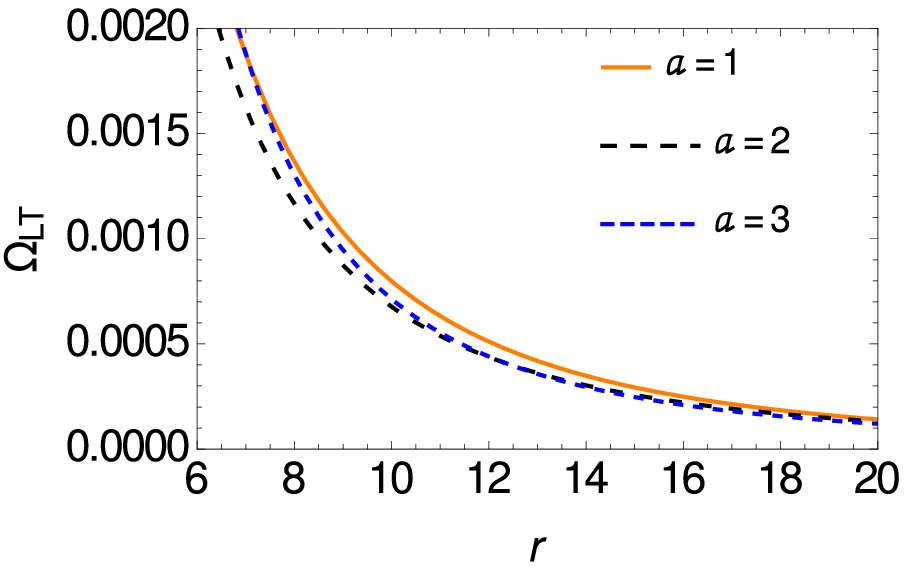}}
 \subfigure[At $30^0$ ]{
 \includegraphics[width=2.3in,angle=0]{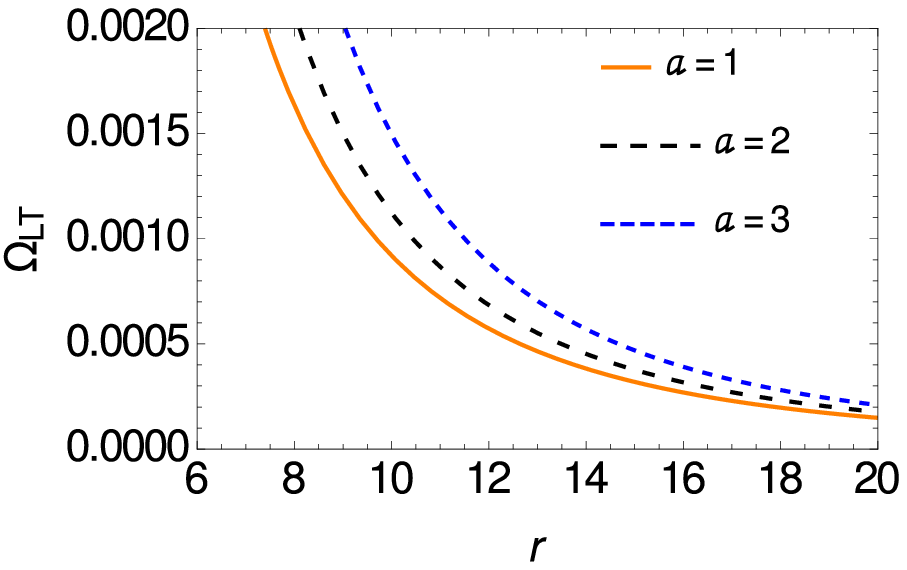}}
 \subfigure[At $90^0$ (along the equator)]{
 \includegraphics[width=2.3in,angle=0]{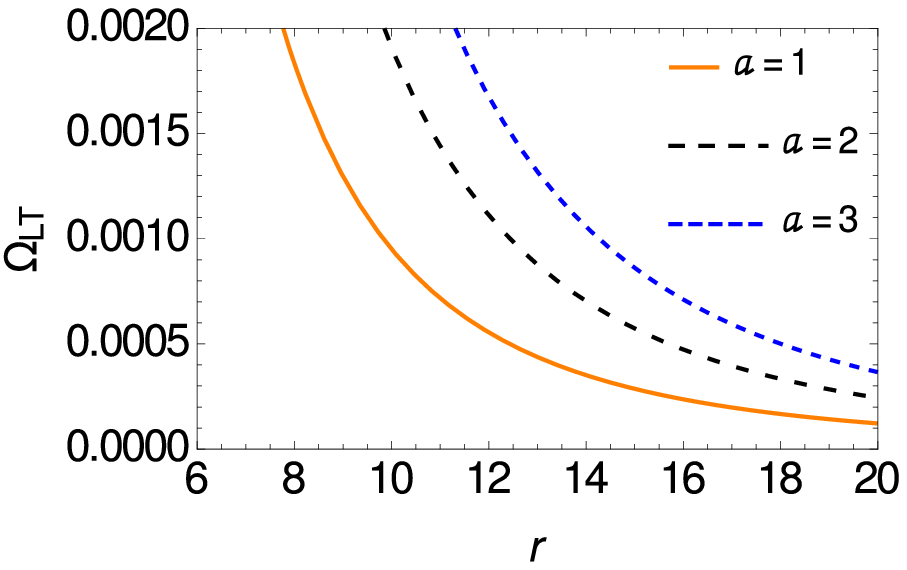}}
\caption{\label{figlt}Radial variation of the LT precession frequency 
($\O_{LT}=|\vec{\Omega}_{LT}|$) in the Teo wormhole for $b=1$ \& $d=1$}
\end{center}
\end{figure}

In Fig.\ref{figlt}, we have plotted  the LT precession frequencies (\ref{sw}) 
of the Teo wormhole to see the 
graphical nature of it. It can be seen from panel (a) of Fig.\ref{figlt} that the LT precession 
frequency $(\O_{\rm LT})$ decreases along the pole with increasing the angular momentum ($a$) of the 
spacetime. This is remarkable in this sense that the LT precession
frequency varies inversely with the rotation of the spacetime 
or so-called the angular momentum of the spacetime. Generally,
frame-dragging rate varies directly with $a$ in case of the black holes
and pulsars but in case of a rotating traversable wormhole its behaviour is slightly
different. The evolution of the frame-dragging rate has been shown in panel (a)-(d) 
of Fig.\ref{figlt}.
If we move from the pole to the equator, the blue line comes upward in a faster 
rate than the orange one and the three colors are slightly indistinguishable in 
panel (b) which is drawn for $\th=12^0$. After crossing the angle $\th \sim 15^0$ with 
respect to the pole, the three colors are 
distinguishable and we can see that the orange line comes down and blue line
comes up which is the general trend as it was generally known to us that if we increases the
rotation of the spacetime
frame-dragging rate increases. In this sense, the nature of the plots of 
Panels (c) and (d) is expected
but the nature of the plots of Panels (a) and (b)
is quite unexpected. Frame-dragging behaves differently only
along the pole and its nearby region ($\sim 10^0$) at the strong gravity
regime in the wormhole spacetime. In case of the weak field limit ($r >> \sqrt{a}$),
it does not behave differently along the pole, we mean, the frame-dragging rate
decreases with decreasing the rotation of the spacetime not only along the pole 
but also along the equator of this wormhole spacetime like the other compact objects
($\O_{LT} \propto a/r^3$) which is evident from Eq.(\ref{sl}) and Eq.(\ref{se}).
In the weak-field limit Eq.(\ref{sl}) and Eq.(\ref{se}) reduce to 
\begin{eqnarray}
  |\O_{LT}|_{\th=0}^{\rm weak} \approx \f{2a}{r^3} 
  \label{ogod}
  \end{eqnarray}
and
\begin{eqnarray}
 |\O_{LT}|_{\th=\pi/2}^{\rm weak} \approx \f{a}{r^3}
 \label{pa}
\end{eqnarray}
respectively, which is expected. Even in the weak gravity regime of the Kerr 
spacetime, LT precession rates 
of a test gyro reduce to $2a/r^3$ along the pole and $a/r^3$ along the equator. 
That is why, $\O_{\rm LT} \propto a$ and $\O_{\rm LT} \propto r^{-3}$
in case of all compact objects as well as in case of the wormholes.
Therefore, the weak-field limit of LT precession cannot distinguish a wormhole 
from the other compact objects but strong field LT precession can do that.
Because, in the strong gravity regime of a rotating wormhole ($r \lesssim \sqrt{a}$),
Eq.(\ref{sl}) and Eq.(\ref{se}) reduce to{\footnote{we have taken $d=1$ 
following Refs.\cite{teo, tb}}}
\begin{eqnarray}
|\O_{\rm LT}|_{\th=0}^{\rm strong} \approx \f{1}{8ar^2}
\label{fp}
\end{eqnarray}
and
\begin{eqnarray}
|\O_{LT}|_{\th=\pi/2}^{\rm strong} \approx \f{3a}{r^3} \left(1-\f{b}{r}\right)^{\f{1}{2}}
\label{fe}
\end{eqnarray}
respectively. It is evident from Eq.(\ref{fp}) that $\O_{LT}$ along the pole is proportional to $a^{-1}$ 
instead of $a$ and this adverse effect nullifies at the distance $r \sim 16a^2$ 
(can be obtained using Eq.(\ref{ogod}) and Eq.(\ref{fp})) along the pole.
It is also interesting to notice that  $\O_{LT}$ is proportional to $r^{-2}$ instead
of $r^{-3}$ along the pole. This anomaly exists only along the pole of the wormhole
and this anomaly vanishes very quickly as we move from the pole to the equator. We can see
from Eq.(\ref{fe}) that no special anomaly arises in the expression of $\O_{LT}$ along the equator
where $\O_{LT}$ varies as $a/r^3$. It is also seen from the plots (Fig.\ref{figlt})
that the value of the LT precession rate along the pole is quite lower than its value
along the equator. Thus, if an astronaut wants to travel through the wormhole,
he/she does not see the higher precession rate of the gyroscope attached to his/her
spaceship (relative to the precession rate along the equator) 
due to the frame-dragging effect, even if the wormhole rotates very fast
as the rapidly rotating wormhole drags its nearby spacetime ($r \sim 16a^2 $) slowly. 

This same anomaly should not be appeared in case of the Kerr black hole or
other axisymmetric spacetimes
like other black holes and pulsars. In the Kerr spacetime, the boundary of 
the outer ergoregion occurs at $r_e=M+\sqrt{M^2-a^2/M^2\cos^2\th}$
where $a$ can take value from $0$ to $M^2$ ($0<a\leq M^2$). Thus, if we take
the limit $r < \sqrt{a}$ (Eq.(42) of \cite{cm}), it will be inside the ergoregion,
which is unphysical and also beyond the frame-dragging
formulation as it is valid only in the timelike region, we mean, outside the ergoregion. 
Thus, we cannot consider the region $r < \sqrt{a}$ which is basically invalid and the region is inside the 
ergosphere. In case of the wormhole, there is no horizon and 
the LT formulation is valid for the range $r > \sqrt{2a\sin\th}$. Now for $\th=0$, 
the LT formulation
is valid in $r \geq b$ along the pole and for $r > \sqrt{2a}$ along the equator and so on. 
Thus, it is meaningful to calculate the LT precession rate for any real
distance equal to the throat radius or greater than this $(r \geq b)$ 
along the pole of the wormhole spacetime. In this situation, if we calculate the LT 
precession rate in the region $b \leq r \sim 16a^2$ using Eq.(\ref{sw}),
we can see that the LT precession
rate decreases with increasing the rotation of the wormhole spacetime but
this treatment is not
applicable for all the angles as $r$ must be greater than $\sqrt{2a\sin\th}$.

It could be noted here that we found an `anomaly' in LT precession in Kerr-Taub-NUT
spacetime \cite{cc2} and inside the pulsars \cite{cmb,ccb} where LT precession does not always follow 
the inverse cube law of distance but the LT precession was proportional to
the rotation of those spacetime everywhere. This is first time when we have found 
a completely different `anomaly' in the LT precession, which is related to
the intrinsic rotation of the spacetime.  

\section{\label{genf}Spin precession of the test gyro in Teo wormhole : general formalism}
In the previous section, we have shown that the spin precession
of a test gyro, which arises due to the LT effect and diverges on the ergoregion. 
This behavior is similar to the cases of Kerr black hole and Kerr naked 
singularity \cite{cm,ckj}. Interestingly, it has recently been shown in \cite{ckj} 
that this behavior is intimately related to and is governed by the geometry 
of the ergoregion in case of a Kerr spacetime. In this regard, we should
discuss the structure of the ergoregion of a Teo wormhole as well.
After that, we will proceed to derive the general formalism of the spin precession
in Teo wormhole so that it can be applied inside as well as outside of
the ergoregion.

\subsection{\label{ergo} Ergoregion in Teo Wormhole and its comparison
with Kerr spacetime}
Teo pointed out that the ergoregion occurs in the range :
$b < r_e \leq \sqrt{2a\sin\th}$ in case of the said wormhole.
According to him, ergoregion could not extend to the poles but it 
forms a tube like structure around the equatorial plane. We find that 
the structure of the ergoregion extends actually in the following angular range : 
 $\sin ^{-1} \f{b^2}{2a} \leq \th_e \leq \pi/2$ and 
  -$\sin ^{-1} \f{b^2}{2a} \geq \th_e \geq -\pi/2$ whereas
  ergoregion absent in the following range :
\begin{eqnarray}
 -\sin ^{-1} \f{b^2}{2a} < \th_{ne} < \sin ^{-1} \f{b^2}{2a}
 \label{ne}
\end{eqnarray}
where $\th_e$ is the angular range which is inside the ergoregion
and $2\th_{ne} (\equiv 2\sin ^{-1} \f{b^2}{2a})$ is the `opening angle', i.e., ergoregion is
absent in this portion.
Therefore, a gyro gets an accessible angular cone around
the polar axis to enter inside the wormhole without touching
the ergoregion. This is necessary for a static observer (whose
angular velocity is zero), which has been considered in Sec.\ref{lt}
to derive the LT precession. For an example, if $a=b=1$,
the ergoregion extends in the following ranges : $\pi/6 \leq \th_e \leq \pi/2$ and 
$-\pi/6 \geq \th_e \geq -\pi/2$ whereas ergoregion absent for
$-\pi/6 < \th_{ne} < \pi/6$. Therefore, the `opening angle' is $\pi/3$ in this example.
However, Eq.(\ref{ne}) reveals that the volume
of the ergoregion increases with increasing the angular momentum
($a$) of the wormhole, i.e., the volume of the ``tube'' around the 
equatorial region increases. For $a \rightarrow \infty$, ergoregion extends
upto $-\pi/2 \leq \th_e \leq \pi/2$ in principle, which means that
the opening angle $(\th_{ne})$ completely vanishes in this case. 
This is in stark contrast with Kerr naked singularity case where it has been shown that
the volume of the ergoregion decreases with increasing the angular momentum 
of the Kerr spacetime (see Eq.(4) of \cite{ckj}) and the volume becomes zero for 
infinite angular momentum, in principle. Interestingly, the ``tube'' like structure
of the ergoregion around the equatorial plane is also formed in case of
a Kerr naked singularity (see FIG. 1 and FIG. 2 of \cite{ckj}), which is quite similar 
to the rotating Teo wormhole case. Only one difference is that,
the radius of the region enclosed by the ``tube'' (throat) is $r=b$ 
(this value does not depend on the value of angular momentum $a$)
at the equatorial plane in case of Teo wormhole whereas this is equal to
the Kerr parameter (radius of the ring singularity) in case of Kerr naked
singularity (see FIG. 2 of \cite{ckj}). 

\subsection{\label{gem} Derivation of the spin precession frequency}
In Sec.\ref{lt}, we have shown that the LT precession does not 
follow the `usual behavior' in the strong gravity regime of a rotating wormhole. 
It could also be noted that the strong gravity LT 
formulation does not valid inside the ergoregion as an observer cannot remain stationary 
there with zero angular velocity. To remain fixed at a particular $r$ and $\th$
inside the ergoregion, the astronaut has to rotate with an angular velocity $\O$ 
within a certain range, so that the velocity of the observer is timelike.
Using this idea, the more general formulation of the spin precession of a test gyro is recently 
developed in Ref. \cite{cpk}. This formulation is valid not only inside
the ergoregion but also outside of it. If a test gyro moves with $\O$, the spin precession 
frequency of it can be calculated using this formulation but we should remember that
this precession does not arise due to only the LT effect. Some additional
precessions are also included with it. We will discuss it as we proceed.
Here, we recapitulate the general formulation of the spin precession which is valid
in any stationary and axisymmetric spacetime and can be expressed as (Eq.14 of \cite{cpk}) 
\begin{eqnarray}
 \vec{\O}_{\rm p}&=&\f{1}{2\sqrt {-g}\left(1+2\O\f{g_{0\phi}}{g_{00}}+\O^2\f{g_{\phi\phi}}{g_{00}}\right)}. \nonumber
 \\
 &&\left[-\sqrt{g_{rr}}\left[\left(g_{0\phi,\th}
-\f{g_{0\phi}}{g_{00}} g_{00,\th}\right)+\O\left(g_{\phi\phi,\th}
-\f{g_{\phi\phi}}{g_{00}} g_{00,\th}\right)+ \O^2 \left(\f{g_{0\phi}}{g_{00}}g_{\phi\phi,\th}
-\f{g_{\phi\phi}}{g_{00}} g_{0\phi,\th}\right) \right]\hat{r} \right. \nonumber
\\
&& \left. +\sqrt{g_{\th\th}}\left[\left(g_{0\phi,r}
-\f{g_{0\phi}}{g_{00}} g_{00,r}\right)+\O\left(g_{\phi\phi,r}
-\f{g_{\phi\phi}}{g_{00}} g_{00,r}\right)+ \O^2 \left(\f{g_{0\phi}}{g_{00}}g_{\phi\phi,r}
-\f{g_{\phi\phi}}{g_{00}} g_{0\phi,r}\right) \right]\hat{\th}\right] . \nonumber
\\
\label{sp}
\end{eqnarray}
It could be noticed from Eq.(\ref{sp}) that $\vec{\O}_{\rm p}$ reduces to 
the LT precession equation ( $\vec{\O}_{\rm p}|_{\O=0}=\vec{\O}_{\rm LT}$)
if $\O$ vanishes. Otherwise, using Eq.(\ref{sp}) we can obtain LT precession 
frequency plus some additional frequencies which arise due to the non-zero
angular velocity of the gyro. For an example, 
LT precession frequency is zero in the Schwarzschild spacetime as it is non-rotating but
the geodetic precession does not vanish in the Schwarzschild spacetime, which arises
due to the non-zero curvature of the spacetime or we can say that it is an effect of the 
non-zero mass ($M \neq 0$) of the spacetime. If a gyro rotates in a circular geodesic
around a Schwarzschild spacetime, $\O$ should be the Kepler frequency :
$\O=(M/r^3)^{\f{1}{2}}$. Thus, the geodetic precession arises due to
the non-zero $M$ or in a broader sense, due to the non-zero angular velocity $\O$. 
Setting $g_{0\phi}=0$ and $\O=$Kepler frequency in Eq.(\ref{sp}), one can obtain
the geodetic precession frequency in a non-rotating spacetime.
This has already been explained in Section IV-C of Ref.\cite{cpk} in detail.
In Kerr spacetime, the total precession frequency is obtained using the full  
expression of Eq.(\ref{sp}) as the LT precession and geodetic precession
both are non-vanishing in this case. However, we can apply Eq.(\ref{sp}) to obtain the 
total spin precession frequency in Teo wormhole but we have to choose a suitable 
range of $\O$ for it. Basically, outside as well as inside of the ergoregion, 
$\O$ can take any value,
provided that the four-velocity ($u=u^{\a}_{\rm obs}=u^t_{\rm obs}~(1,0,0,\O)$) 
of the gyro be timelike. Therefore, the range of $\O$ is calculated as \cite{cpk}:
\begin{eqnarray}
 \O_- (r,\th) < \O < \O_+ (r,\th)
\end{eqnarray}
where,
\begin{eqnarray}
 \O_{\pm}=\o \pm \f{1}{r \sin \th} =\f{2a}{r^3} \pm \f{1}{r\sin\th}
 \label{o}
\end{eqnarray}
in case of the Teo wormhole. Otherwise, the gyro cannot remain
stationary inside of the ergoregion.
For simplicity, we introduce a parameter $q$ to scan the allowed values for $\O$ 
 inside the ergoregion ($b < r \leq \sqrt{2a\sin \th}$) of the said wormhole,
 which follows
 \begin{eqnarray}
 \O &=& q~\O_+ + (1-q)~\O_-=\O=\o + \f{(2q-1)}{r \sin \th} 
 \label{o}
\end{eqnarray}
where $0 < q <1$. Now, in case of a Teo wormhole, we can express
the total spin precession frequency of a test gyro (which rotates with an 
arbitrary $\O$) as
\begin{eqnarray} \nonumber
 \vec{\Omega}_{p}&=&\frac{1}{4q(1-q)~Nr^3}.
\left[(2a-\O~r^3)\cos\theta ~\hat{r} \right.
\\
&& \left. + N\sin\theta\left(1-\frac{b}{r}\right)^{\frac{1}{2}}
\left\{ a\left(1+\frac{12a^2}{r^4}\sin^2\theta \right)
+\O r^3 \left(1-\frac{12a^2}{r^4}\sin^2\theta \right)+3ar^2\O^2 \sin^2\th \right\} \hat{\theta}
\right] \nonumber .
\\ 
\label{spin}
\end{eqnarray}
We note that the above expression is valid for $r \geq b$ and $ 0 < \th \leq \pi/2$. 
Teo wormhole is a {\it simplest} possible rotating traversable wormhole
in this sense that only the throat radius $b$ and angular momentum $a$ have been introduced in it.
Therefore, it is possible to identify the effect of these two parameters 
in the spin precession,
which is quite interesting. Teo wormhole is not explicitly depend on the mass
parameter of the spacetime, which is basically responsible for the geodetic
precession in Schwarzschild and Kerr spacetimes. Eq.(\ref{spin}) reveals that 
the spin precession arises due to the frame-dragging effect as well as the 
non-zero rotation ($\O$) of the spin. As a special case, if the gyro rotates in a circular 
geodesic in the Schwarzschild spacetime, $\O$ arises due to the non-zero mass.
Thus, $\O_{\rm p}$ gives the geodetic precession frequency.
Now, if the gyro rotates in a circular geodesic of a {\it non-rotating} ($a=0$) Teo wormhole, 
it can easily be shown using Eq.(\ref{kd}) and Eq.(\ref{kg}) that $\O_{p}$ vanishes 
as Kepler frequency $(\propto a/r^3)$ depends solely on the angular momentum
($a$) of the wormhole. We argue that as the Teo wormhole or the Kepler frequency
has no explicit mass dependence, geodetic precession does not arise in this case.
It is needless to say here that LT precession is zero as the spacetime is 
non-rotating. Therefore, an astronaut does not see any precession
of the test spin/gyro attached to his/her spaceship in this case.
This interesting phenomena is completely absent 
in the cases of other compact objects. If the gyro does not move along
a geodesic (as it pointed out earlier in Refs. \cite{hoj,abk} that a gyro does 
not move along the geodesic in strong gravity), $\O$ may not be zero
even in case of $a=0$. It can be seen from Eq.(\ref{o}) that the angular velocity 
of the stationary observer $\O$ reduces to $\O=(2q-1)/(r\sin\th)$ when
it moves toward a non-rotating wormhole ($a=0$) spacetime.
Therefore, the spin precession arises due to only the rotation of the test gyro
in this case. We are not interested in the non-rotating wormhole
in our present article but what we want to emphasize it here
is that the spin precession not only arises due to the curvature 
of the spacetime or rotation of the spacetime but it also arises
due to the non-zero angular velocity of the spin when it does not move along a
geodesic in the strong gravity regime. This special effect
is absent if the gyro moves along a geodesic. 

\begin{figure}[]
 \begin{center}
  \subfigure[For three different values of $q$ at $\th=60^0$ with $a=1$]{
 \includegraphics[width=2.7in,angle=0]{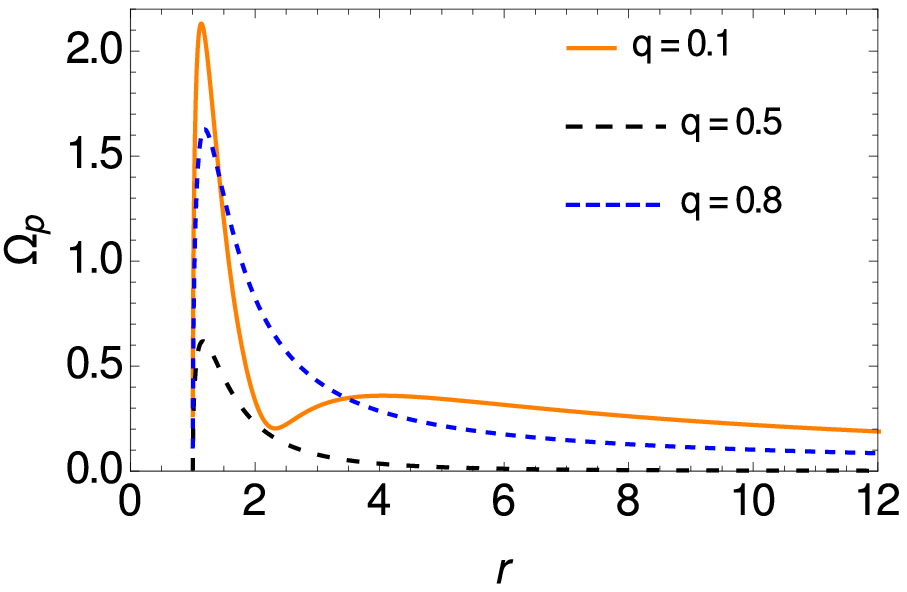}}
 \subfigure[In three different angles with $a=1$ and $q=0.3$]{
 \includegraphics[width=2.7in,angle=0]{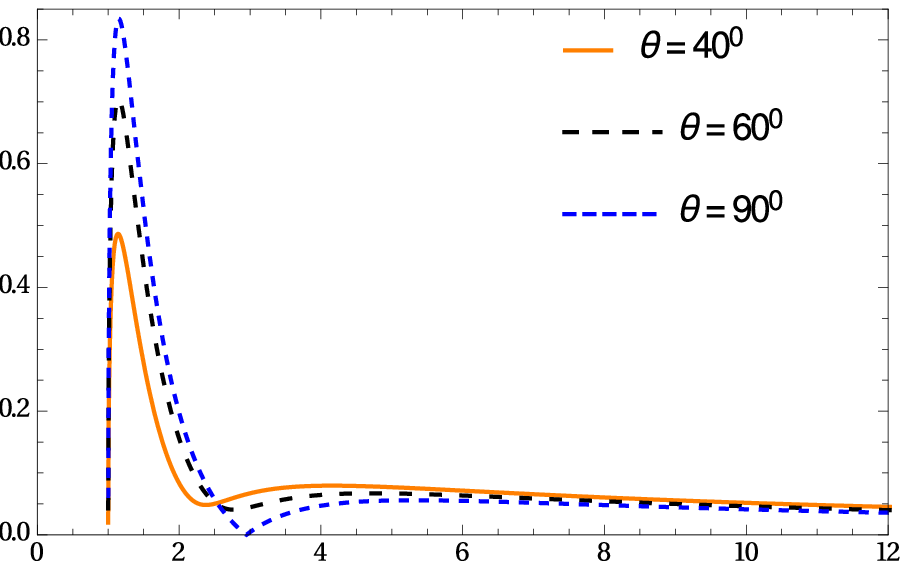}}
 \subfigure[For three different values of $a$ at $\th=60^0$ with $q=0.3$]{
\includegraphics[width=2.7in,angle=0]{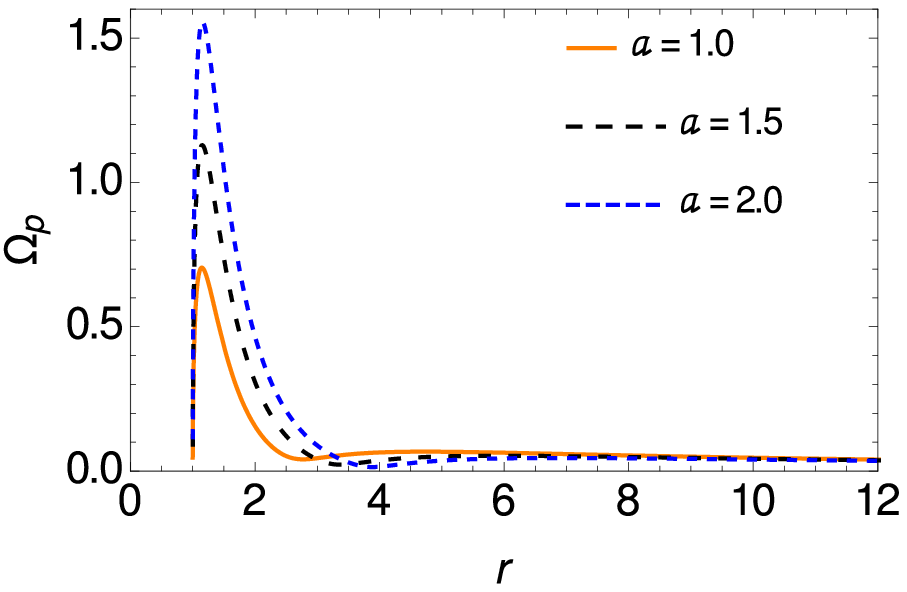}} 
\caption{\label{figsp}Plot of spin precession frequency ($\O_{\rm p}$)
of a test gyro vs distance ($r$) for three different circumstances
in the wormhole spacetimes for $b=1$ \& $d=1$}
\end{center}
\end{figure}

In Fig.\ref{figsp} we plot the radial variation of the spin 
precession frequency $\O_{\rm p}=|\vec{\Omega}_{p}|$ to see the nature
of the spin precession in Teo wormhole. The ergoregion extends
from $b < r_e \leq \sqrt{2a\sin\th}$ in all the plots where 
we have taken $b=1$ and the maximum and minimum values of $a$ are
taken as $a=2$ and $a=1$ respectively. Thus, the ergoregion 
occurs in the following ranges : $1 < r_e \leq 1.86$ for $a=2$ and 
$1 < r_e \leq 1.31$ for $a=1$ at $\th=60^0$. Panel (a) shows the
radial variation of $\O_{\rm p}$ for three different values of $q$. 
It is clearly seen from this plot that a local maxima and a local
minima appear for $q=0.1$. Actually, this anomaly occurs for 
any value of $q : 0 < q < 0.5$ but it is absent for $0.5 \leq q < 1$.
If $q$ increases from $0$ to $0.5$, the value of $\O_{\rm p}$ decreases 
and the local maxima ($r_x$) and minima ($r_n$) disappear for $q=0.5$. Further 
increment of $q$ from $0.5$ to $1$, the value of $\O_{\rm p}$ increases 
but the local maxima and minima do not appear again.
Panel (b) shows that the spin precession rate increases at 
$r=r_x$ and it decreases for $r \geq r_n$,
if the gyro moves from the pole to the equator of the wormhole.
Panel (c) shows that $\O_{\rm p}$ is higher for higher $a$ at the local
maxima but it is completely opposite for $r=r_n$. The later feature
is quite similar to Panel (a) and Panel (b) of Fig.\ref{figlt}. We note that
the spin precession frequency at the throat of rotating Teo wormhole is
\begin{eqnarray}
 \vec{\Omega}_{p}|_{r\rightarrow b} \rightarrow 
 \frac{(1-2q)\cot \th ~\hat{b} }{4q(1-q)~(b+16a^2d\cos^2\th)}
 \label{b}
\end{eqnarray}
which vanishes for $q=0.5$ (see Panel (a) of Fig.\ref{figsp}) or $\th=\pi/2$.
It could be noticed from Eq.(\ref{b}) that the spin precession does not 
vanish even if $a=0$ but it vanishes for $q=0.5$ as in that case $\O_p$ arises
due to frame-dragging effect only. We discuss it vividly in the next section.

\subsection{Special Case : ZAMO}
$q=0.5$ is a very special case which is also quite interesting
in general relativity as well as in Astrophysics. For $q=0.5$, 
Eq.(\ref{o}) reduces to 
\begin{eqnarray}
 \O=\o=\f{2a}{r^3}.
\end{eqnarray}
Only in this case, the test gyro can regard the $+\phi$ and $-\phi$
directions as equivalent in terms of the local geometry. This can also
be said as the gyro is ``non-rotating relative to the local 
spacetime geometry'' or we can say it as ``locally non-rotating observer/gyro''
who moves with the angular velocity $\O=\o$. Thus, the angular momentum 
of this ``locally non-rotating observer'' is zero, for which the observer is called
as the Zero-Angular-Momentum-Observer (ZAMO) which was first introduced
by Bardeen \cite{bd,mtw}. Bardeen et al.\cite{bpt} showed that 
the ZAMO could be a powerful tool in the analysis of physical
processes near astrophysical object. It is needless to say that 
our general expression (Eq.\ref{sp}) of spin precession is valid not only for 
$q=0.5$ but also for all possible values of the angular 
momentum ($\O$) of the gyro, i.e., $0 < q < 1$. However,
as it has been described in Refs.\cite{bs,mtw} that the angular velocity
of the photon could be expressed by $\o$, one can calculate in principle the 
spin precession frequency using our general expression Eq.(\ref{spin}).
This can be written as :
\begin{eqnarray}
 \vec{\Omega}_{p}|_{q=1/2}=\frac{3a\sin\th\left(1-\frac{b}{r}\right)^{\frac{1}{2}}}{r^3}~\hat{\th}
. \label{ph}
 \end{eqnarray}
The above expression (Eq.\ref{ph}) reveals that the spin precession 
frequency is directly proportional to $a$, which means that the 
spin precession arises in this case due to only the frame-dragging effect,
i.e., the non-zero rotation of the spacetime. In some articles,
ZAMO has been introduced to derive the frame-dragging effect in this special situation
(for an example, see Ref.\cite{ms} where the frame-dragging frequency
was derived for rapidly rotating neutron stars using ZAMO). 
Earlier, we derived the radius of CPO
at the equatorial plane in Sec.\ref{cpow}, which came out as
$r_c \approx 2.4626\sqrt{a}$ (Eq.\ref{cpo}). Now, substituting this 
value in Eq.(\ref{ph}) we obtain the spin precession frequency of 
the equatorial photon as
\begin{eqnarray}
 \vec{\Omega}_{p}|_{q=1/2,\th=\pi/2}
 =\pm \frac{0.2}{\sqrt{a}}\left(1-\frac{b}{2.4626\sqrt{a}}\right)^\frac{1}{2}~\hat{\th}
. \label{phlt}
 \end{eqnarray}
It indicates that if the throat radius of the wormhole is equal to 
$2.4626\sqrt{a}$, spin precession vanishes at the throat though 
it is located in strong gravity regime. This does not arise for 
$b > 2.4626\sqrt{a}$ because, the precession frequency becomes imaginary  
and CPO occurs inside the throat but
we can always calculate the spin precession of the photon in the opposite
case, i.e., $b \leq 2.4626\sqrt{a}$, as CPO occurs outside the throat 
of a Teo wormhole.

\section{\label{dis}Summary and Discussion}
In the first part of this article, we have briefly discussed the geodesic 
structure of the rotating traversable Teo wormhole. 
Later, we have derived some observables which might be used to detect a rotating 
wormhole, if it in fact exists.
Namely, we have indirectly shown that the ISCO coincides with the throat in Teo wormhole 
for the prograde rotation, as the effective potential vanishes
due to $\ell_{0}=0$ and ${\cal E}_{0}=1$ in this case whereas for the 
retrograde rotation, only one stable circular orbit exists 
which occurs at the throat. After that, we have derived the fundamental frequencies
of a test particle which rotates in an  equatorial circular orbit in this
rotating wormhole.
We have shown that the Kepler frequency $(\O_{\phi})$ is
proportional to $r^{-3}$ in the Teo wormhole but in general it varies with
$r^{-\f{3}{2}}$ in case of the Kerr
black hole and other similar compact objects like pulsars, neutron stars etc.
It has also been shown that the Kepler frequency is directly proportional to 
the angular momentum of the wormhole. Therefore, it vanishes in case 
of the non-rotating Teo wormhole. Similarly, the nodal plane precession 
frequency and periastron precession frequency are also proportional to $r^{-3}$ 
in this case. This can be used to distinguish between a Kerr spacetime and a 
wormhole spacetime, if the latter in fact exists.
Interestingly, the Kepler frequency and the periastron precession frequency
behave in a similar fashion due to the vanishing radial epicyclic frequency for the prograde rotation.
Thus, we can conclude that the Kepler frequency and periastron precession frequency
are actually same in a Teo wormhole. 

In the second part, we have shown that the LT
precession frequency of a test gyro need not always be directly proportional to
the rotation of the wormhole. It behaves differently in the wormhole
spacetime. In fact, we have find that the LT precession rate
along the pole is inversely proportional to the 
angular momentum of the wormhole, contrary to our usual
expectation that the LT precession rate is not only directly proportional to 
the rotation of the spacetime but also follows the inverse cube law of distance.
Here, we have deduced that the LT precession rate follows an inverse square law 
of distance along the pole in the strong gravity regime.
We believe these to be the crucial factors in observationally distinguishing a wormhole
spacetime from that of other compact objects as such an
`adverse effect' has not been found in any other spacetimes till now.
In our recent study on  LT precession in a Kerr 
spacetime \cite{ckj}, we have demonstrated that the LT precession frequency is never 
 inversely proportional to 
the rotation of the spacetime outside a Kerr black hole because it should always be
proportional to the angular momentum of the spacetime though LT precession acts differently in case of
a Kerr naked singularity. 
Therefore, the distinct behaviors of  LT precession may help us to distinguish 
between wormholes, Kerr black holes and Kerr naked singularities during future
astrophysical observation. Further, using the general formulation of spin precession
\cite{cpk}, we have derived the spin precession frequency of a test gyro inside and
outside  the ergoregion  of a Teo wormhole. It has been shown that 
the spin of a gyro precesses not only due to the curvature and the rotation
of the spacetime but also due to the non-zero angular velocity of the gyro.
This general formulation reduces to the case of a ZAMO for $q=0.5$
when the spin precession arises purely due to the frame-dragging effect.
We have also clearly shown, at least in principle, that there are key and
essential theoretical differences that can be used to distinguish a 
wormhole from a black hole. 
Finally, we can say that if in  future, interstellar traveling through a wormhole 
 or at least in its vicinity becomes possible, our results would give detailed information on 
the behavior of a test gyroscope which  serves  as a fundamental 
navigation device.
\\

{\bf Acknowledgements :} We thank Oindrila Ganguly for her careful and critical reading
of the manuscript.

\end{document}